\documentclass[fleqn,twoside]{article}
\usepackage[headings]{espcrc2}

\readRCS
$Id: espcrc2.tex,v 1.2 2004/02/24 11:22:11 spepping Exp $
\usepackage{graphicx}
\usepackage{amsmath}
\usepackage{bm}

\title{Recent Developments in Heavy-Quarkonium Phenomenology}

\author{Hee Sok Chung\address[KU]{Department of Physics, Korea University, 
        Seoul 136-701, Korea},
        Jungil Lee\addressmark[KU]\thanks{This work was supported by the Korea 
        Science and Engineering Foundation (KOSEF) funded by the Korea government
        (MEST) under grant No. R01-2008-000-10378-0.}
        and Chaehyun Yu\addressmark[KU]}
       
\runtitle{Recent Developments in Heavy Quarkonium Phenomenology}
\runauthor{Hee Sok Chung, Jungil Lee and Chaehyun Yu}

\begin{document}

\begin{abstract}
We review recent developments in heavy-quarkonium phenomenology within
the nonrelativistic QCD factorization approach. Main issues we consider
in this work include the polarization of prompt $J/\psi$ at the Fermilab
Tevatron and the large relativistic and QCD corrections to
double-charmonium production at the $B$ factories. We also consider
inclusive charm production in bottomonia decays.
\vspace{1pc}
\end{abstract}

\maketitle
\section{Introduction\label{sec:intro}}
For more than a decade, the nonrelativistic quantum
chromodynamics (NRQCD) factorization approach~\cite{Bodwin:1994jh}
has been employed to predict various measurables in
heavy-quarkonium phenomenology. In this approach, rates
for production or decay of a heavy quarkonium that involve creation
or annihilation of a heavy-quark-antiquark ($Q\bar{Q}$) pair are
expressed in linear combinations of NRQCD matrix elements, which
reflect the long-distance nature of the heavy quarkonium.
Corresponding short-distance coefficients are insensitive to the
long-distance interaction and calculable perturbatively. As the
first application of the factorization theorem for the decays,
the light-hadronic and electromagnetic decay rates of the $P$-wave
heavy quarkonia were factorized~\cite{Bodwin:1992ye}. Explicit
infrared singularities appearing in order-$\alpha_s^3$ 
color-singlet-model predictions for the light-hadronic decays of
$P$-wave heavy quarkonia~\cite{Barbieri:1976fp,Barbieri:1980yp}
became to be identified as those of the next-to-leading-order (NLO)
corrections to the long-distance color-octet NRQCD matrix elements
that mix with the color-singlet NRQCD matrix elements at this
order.

Soon after the resolution of the infrared problem in the
$P$-wave quarkonium decay,  $\psi(2S)$ anomaly was reported:
the transverse-momentum $(p_T)$ distributions for prompt $\psi(2S)$
and $J/\psi$ measured by the CDF Collaboration at the Fermilab Tevatron
were greater than theoretical predictions of the color-singlet
model by orders of magnitude at large
$p_T$~\cite{Abe:1997jz,Abe:1997yz}.
Based on the idea of gluon-fragmentation dominance
of quarkonium production at large $p_T$~\cite{Braaten:1993rw},
the puzzle was resolved by introducing the color-octet $S$-wave
spin-triplet [$Q\bar{Q}_8({}^3S_1)$] contribution to the gluon
fragmentation into an $S$-wave spin-triplet
heavy quarkonium~\cite{Braaten:1994vv}.
The dominance of the color-octet mechanism in large-$p_T$
prompt $\psi(2S)$ production led to the prediction that
the prompt $\psi(2S)$ must be transversely polarized
at large $p_T$~\cite{Cho:1994gb,Cho:1994ih}. Unfortunately,
the predictions of the polarizations of prompt
$\psi(2S)$~\cite{Beneke:1995yb,Leibovich:1996pa}
and prompt $J/\psi$~\cite{Braaten:1999qk,Kniehl:2000nn} 
were disfavored by the CDF measurement at Run I of the
Fermilab Tevatron~\cite{Affolder:2000nn}. For more details,
we refer the readers to 
Refs.~\cite{Brambilla:2004wf,Bedjidian:2004gd}. 
This problem has been a serious challenge to 
heavy-quarkonium theory. 

Another interesting problem in quarkonium phenomenology
that has not been resolved until very recently is
regarding the cross section for
the exclusive $J/\psi+\eta_c$ production in $e^+e^-$
annihilation at $B$ factories. The cross section
measured by the Belle Collaboration~\cite{Abe:2002rb}
was greater than theoretical predictions at
leading order (LO) in $\alpha_s$ and $v$
by an order
of magnitude~\cite{Braaten:2002fi,Liu:2002wq}.
The BABAR Collaboration also observed a large
cross section~\cite{Aubert:2005tj} and
subsequent studies~\cite{Abe:2004ww,Yabsley:2007jb} 
of the Belle Collaboration revealed
that the large discrepancy between theory and data is not
originated from the contamination of other
signals~\cite{Bodwin:2002fk,Bodwin:2002kk,Brodsky:2003hv}.
On the other hand, there were arguments that an appropriate
choice of light-cone distribution amplitude may resolve the
problem~\cite{Bondar:2004sv,Ma:2004qf}, which were confronted with
an opposing argument that, after separating the short-distance
contribution from the long-distance light-cone distribution
amplitude, the light-cone method yields a result equivalent
to that of NRQCD~\cite{Bodwin:2006dm}.
Very recently, large relativistic
corrections~\cite{Bodwin:2006dn,Bodwin:2006ke}
in combination with one-loop QCD
corrections~\cite{Zhang:2005cha,Gong:2007db}
filled the gap within errors~\cite{Bodwin:2007ga}.

In this work we review recent developments in heavy-quarkonium
phenomenology within the nonrelativistic QCD factorization approach,
focusing on the following three selected topics.
Issues regarding the polarization of prompt $J/\psi$ at the Fermilab
Tevatron and future opportunities to resolve the problem at the CERN LHC are
presented in Sec.~\ref{sec:pol}. The discrepancy between the
LO predictions and the empirical cross sections for 
$e^+ e^- \to J/\psi + \eta_c$ measured at the $B$ factories has been
a long-standing puzzle. In Sec.~\ref{sec:rel} we briefly review the
problem and describe a recently developed method of resumming a class
of relativistic corrections to all orders in the heavy-quark velocity
$v$, which played a crucial role in resolving the 
$e^+ e^- \to J/\psi + \eta_c$ problem in conjunction with the 
relative-order-$\alpha_s$ corrections. As the last topic,
inclusive charm productions in bottomonia decay being analyzed by
the CLEO Collaboration and corresponding theoretical predictions are
reviewed in Sec.~\ref{sec:bottom} which is followed by conclusions in 
Sec.~\ref{sec:sum}.
\section{Polarization of prompt $\bm{J/\psi}$\label{sec:pol}}
The discrepancy between the theoretical prediction and the CDF measurement
at Run I of the Fermilab Tevatron for the prompt $S$-wave spin-triplet
charmonium produced at large $p_T$ was embarrassing. Gluon fragmentation
has the kinematic
enhancement factor of $p_T^2/m_{J/\psi}^2$ which dominates over the
suppression factor $\alpha_s^2(p_T)$ for the coupling compared to the
gluon-fusion diagrams at sufficiently large $p_T$~\cite{Braaten:1993rw}.
In comparison with the gluon fragmentation into the color-singlet
$Q\bar{Q}$ pair [$Q\bar{Q}_1({}^3S_1)$], the 
$Q\bar{Q}_8({}^3S_1)$ contribution has a less
severe suppression factor of order $v^4$ which accounts for
the spin-non-flipping long-distance transition from the color-octet
to the color-singlet state, while the suppression factor
$\alpha_s^2(p_T)$ in the color-singlet fragmentation channel 
is for the short-distance coefficient~\cite{Braaten:1994vv}.
If the gluon fragmentation into the $Q\bar{Q}_8({}^3S_1)$ pair is
the dominant contribution to the prompt $J/\psi$ and $\psi(2S)$ 
production at
large $p_T$, then the invariant mass for the fragmenting virtual gluon
is negligible compared to the typical scale $p_T$ of the short-distance
process. Hence, longitudinally polarized virtual-gluon contribution
must be suppressed compared to the transverse contribution with a
multiplicative scaling factor of
$m_{J/\psi}^2/p_T^2$~\cite{Cho:1994gb,Cho:1994ih}. Because the long-distance
transition from $Q\bar{Q}_8({}^3S_1)$ to $Q\bar{Q}_1({}^3S_1)$ is 
dominated by spin-non-flipping chromoelectric dipole transitions,
prompt $\psi(2S)$ must be transversely polarized at large $p_T$.
This is also supported by a lattice computation of the color-octet
NRQCD matrix element involving the spin-flipping
interaction for the decay~\cite{Bodwin:2005gg}.
On the contrary, the predictions for the $\psi(2S)$
polarization~\cite{Beneke:1995yb,Leibovich:1996pa,Braaten:1999qk} were
disfavored by the CDF data at both Run I and Run II of the Fermilab
Tevatron~\cite{Affolder:2000nn,Abulencia:2007us}.

In a similar manner as $\psi(2S)$, one can expect that the prompt
$J/\psi$, which has much larger statistics than $\psi(2S)$,
produced at large $p_T$ must be transversely polarized.
However, prompt
$J/\psi$ suffers from feeddowns from higher resonances such as
$\psi(2S)$ and $\chi_{cJ}$. At the Tevatron, the feeddowns from
$\psi(2S)$ and $\chi_{cJ}$ are about $10$\% and $30$\%, respectively.
The contribution from $\chi_{cJ}$ dilutes the polarization and
that from $\psi(2S)$ is dominantly spin
preserving~\cite{Braaten:1999qk,Kniehl:2000nn}.
Although feeddown from higher resonances makes the prediction for
the prompt $J/\psi$ less dramatic than $\psi(2S)$, the tendency of
rising transverse polarization as $p_T$ increases is the same as
that for $\psi(2S)$~\cite{Braaten:1999qk,Kniehl:2000nn}. Again,
this expectation was confronted with the CDF data at both Run I
and Run II of the Fermilab 
Tevatron~\cite{Affolder:2000nn,Abulencia:2007us}.
One may suspect any flaws in the theoretical
predictions. The prediction for the prompt $J/\psi$
polarization in Ref.~\cite{Braaten:1999qk,Kniehl:2000nn} was
at LO in $\alpha_s$ and $v$. One may expect that
the inclusion of real gluon emission and higher-order
contributions in $v$ that flip spin may dilute the polarization
even if they are suppressed based the velocity-scaling
rules of NRQCD.
One may also suspect that the dominance of gluon fragmentation
has not yet been achieved in the $p_T$ range accessible 
at the Fermilab Tevatron.
At the same time, it is fair to notice that the discrepancy between
the Run I and Run II data of the CDF Collaboration is not
clearly understood, yet~\cite{Affolder:2000nn,Abulencia:2007us}.

Theoretical prediction for the $\Upsilon(nS)$ produced at the 
Fermilab Tevatron
was not dramatic as the $S$-wave spin-triplet charmonium state because the
available range of $p_T$ 
does not reach the region where gluon fragmentation
dominates~\cite{Braaten:2000gw}.
Theoretical uncertainties were large because of the large errors
in various NRQCD matrix elements of bottomonia~\cite{Braaten:2000cm}. 
The CDF measurement of the $\Upsilon(nS)$ polarization~\cite{Acosta:2001gv}
was consistent with the prediction within errors~\cite{Braaten:2000gw}.
Around the time when the prompt $J/\psi$ polarization measurement of 
the CDF Collaboration was announced, the E866 Collaboration reported that
$\Upsilon(1S)$ is almost unpolarized while the samples for
$\Upsilon(2S)+\Upsilon(3S)$
are almost transversely polarized~\cite{Brown:2000bz} in their fixed-target
experiment. This shows that the polarization of $\Upsilon(1S)$, that has
many channels of feeddown from higher resonances through $P$-wave states,
is diluted compared to $\Upsilon(2S)$ and $\Upsilon(3S)$ states.
Similar tendency has been observed by the D0 Collaboration at 
Run II of the Tevatron~\cite{:2008za}. The D0 result for $\Upsilon(1S)$
shows that $\Upsilon(1S)$ is longitudinally polarized at most of
$p_T$ regions below about $15~$GeV. This may be due to the feeddown effect
through $P$-wave resonances. However, transverse contribution
increases continuously from the maximally longitudinal region around
$2$--$5$~GeV as $p_T$ increases.  The D0 result for $\Upsilon(2S)$, which has
less feeddown effect than $\Upsilon(1S)$ shows that the rise of transverse
polarization is faster and stronger than that of $\Upsilon(1S)$.
This indicates that polarization analysis for $\Upsilon(nS)$ at the
CERN LHC, where accessible $p_T$ may be large enough to approach
gluon-fragmentation-dominance region for $\Upsilon(nS)$ production,
may reveal transversely polarized spin-triplet $S$-wave bottomonia.
In comparison with the charmonium counterparts, $\Upsilon(nS)$ is
more nonrelativistic. As a result, it will suffer less from the 
contamination of longitudinal contributions of
higher orders in $\alpha_s$ and $v$.

We list recent theoretical attempts that have been made
to improve the NRQCD predictions
for the inclusive heavy-quarkonium production at hadron colliders.
The QCD NLO corrections to the color-singlet channel to the
gluon-fusion diagrams, $gg \to Q\bar{Q}_1(^3S_1) + g$, 
that contributes to the ${}^3S_1$ quarkonium production at
hadron colliders were recently 
computed~\cite{Campbell:2007ws}. It was reported that
the NLO corrections are large and modify the expectations for 
the total cross sections and the distribution in the transverse
momentum. This is a signal that the NRQCD matrix element for the
long-distance process $Q\bar{Q}_8(^3S_1)\to J/\psi+X$ may be smaller
than that estimated before, resulting in less dramatic increase of
transverse polarization of prompt $J/\psi$. 
In Ref.~\cite{Artoisenet:2007xi},
the contribution of the following specific channels,
$g g \to Q\bar{Q}_1(^3S_1) + Q\bar{Q}$,
where $Q$ is $c$ or $b$,  were calculated. The authors reported that,
in those channels, fragmentation approximation underestimates
the cross section in the kinematical region accessible at the Fermilab Tevatron.
In addition, the polarization of both $J/\psi$ and $\Upsilon$ 
produced through above channels are unpolarized.
In Ref.~\cite{Artoisenet:2008fc}, the authors update the theoretical
predictions for hadroproduction of direct $\Upsilon$. They reported
that the QCD NLO corrections to the color-singlet 
contribution significantly enhances
the theoretical prediction for the cross section at large $p_T$
and substantially affect the polarization of the
$\Upsilon$~\cite{Artoisenet:2008fc}.
Because the measured polarizations are for the inclusive
$\Upsilon$ that contain various feeddowns, one cannot 
make a direct comparison with the available data for the inclusive
$\Upsilon(nS)$. Further theoretical predictions for
partial contributions to the inclusive charmonium production
at hadron colliders are also
available~\cite{Gong:2008sn,Gong:2008hk,Gong:2008ft}.
\section{Relativistic corrections to quarkonium processes\label{sec:rel}}
In general, because of clean initial states, processes for quarkonium
production in $e^+e^-$ annihilation may have less theoretical uncertainties
than hadroproduction. Especially, the exclusive double-charmonium
production from $e^+e^-$ annihilation into a virtual photon should be
described relatively with ease because the color-singlet contribution in NRQCD
is absolutely the dominant production mechanism. For the exclusive process
$e^+e^-\to J/\psi+\eta_c$, only a single long-distance
color-singlet NRQCD matrix element, $\langle \mathcal{O}_1\rangle_{J/\psi}$,
involves, once we impose the approximate heavy-quark spin symmetry
of NRQCD. $\langle \mathcal{O}_1\rangle_{J/\psi}$
is the most accurately known NRQCD matrix element, which is 
determined from $\Gamma[J/\psi\to e^+e^-]$.

However, the cross section first measured by the Belle Collaboration, 
which is about $33$ fb~\cite{Abe:2002rb}, was
greater than the NRQCD predictions~\cite{Braaten:2002fi,Liu:2002wq} 
at LO in both $\alpha_s$ and $v$
by an order of magnitude.
Later the Belle Collaboration updated the value as
$25.6\pm 2.8\pm 3.4$ fb~\cite{Abe:2004ww} and the BABAR Collaboration
also reported a large cross section of
$17.6\pm 2.8\pm 2.1$ fb~\cite{Aubert:2005tj},
where the values include the branching fraction of the system recoiling
against the $J/\psi$ into two or more charged tracks.
The theoretical predictions were 
$3.78\pm 1.26$ fb~\cite{Braaten:2002fi} and 
$5.5$ fb~\cite{Liu:2002wq}, where the difference depends on the choice
of input parameters. The prediction in Ref.~\cite{Braaten:2002fi}
also includes QED contributions that make a constructive interference
of about $30$\% of the total cross section for the QCD process.
As was described earlier, proposals that
the signal may include final states other than 
$J/\psi+\eta_c$~\cite{Bodwin:2002fk,Bodwin:2002kk,Brodsky:2003hv} were
disfavored by an updated analysis by the Belle 
Collaboration~\cite{Abe:2004ww,Yabsley:2007jb}.
And the light-cone method reproduced the NRQCD
prediction~\cite{Bodwin:2006dm}.
This puzzle has been a serious challenge to quarkonium
phenomenology. 

When the first theoretical prediction was made,
it was reported that there may be huge relativistic corrections
to the LO prediction~\cite{Braaten:2002fi}.
Unfortunately, there was no reliable
way available to estimate  the
NRQCD matrix element involving relativistic corrections, which is
power-ultraviolet-divergent and needs subtraction~\cite{Braaten:2002fi}.
This was an obstacle to evaluating relativistic corrections to the process.
As the next step, QCD NLO corrections
to the $e^+e^-\to J/\psi+\eta_c$ process were calculated. 
The $K$ factor for the QCD NLO prediction to that of 
LO in both $\alpha_s$ and $v$ was about a factor of
2~\cite{Zhang:2005cha,Gong:2007db}.
In spite of a large enhancement from the QCD NLO corrections, 
large discrepancy between theory and experiments still remained.

The only sector that had not been explored thoroughly was relativistic
corrections to the process. The first prediction of
the $K$ factor for the relativistic corrections was 
$2.0^{+10.9}_{-1.1}$~\cite{Braaten:2002fi}. The main source for the
huge uncertainty was from the relative-order-$v^2$ NRQCD matrix element.
In principle, once the quarkonium color-singlet wave function is known
by using a non-perturbative method such as lattice 
NRQCD~\cite{Bodwin:1996tg,Bodwin:2001mk}, one can compute
the order-$v^2$ NRQCD matrix element, where power-ultraviolet-divergent
contributions to the matrix element must be subtracted with dimensional
regularization to make it consistent with the standard NRQCD factorization
formalism. Because of this large subtraction,
even the sign of the matrix element was uncertain by the time the
prediction was made~\cite{Braaten:2002fi}. 

As a breakthrough, a new method to determine the
color-singlet 
$S$-wave NRQCD matrix elements was introduced~\cite{Bodwin:2006dn}.
 Employing the Cornell potential~\cite{Eichten:1978tg}, which is
 a linear combination of
the Coulomb potential and the linear potential, the authors computed the
quarkonium color-singlet wave function~\cite{Bodwin:2006dn,Chung:2008sm}.
Relative strength of the Coulomb potential compared
to the linear potential was determined by making use of the lattice 
string tension~\cite{Bodwin:2006dn}.
Furthermore, they derived the relation, 
$\langle \bm{q}^{2n} \rangle_H=\langle \bm{q}^2 \rangle_H^n$~\cite{Bodwin:2006dn},
which is the generalization of the Gremm-Kapustin
relation~\cite{Gremm:1997dq}, where $\langle \bm{q}^{2n}\rangle$ is
the ratio of the relative-order-$v^{2n}$
color-singlet matrix element to 
$\langle \mathcal{O}_1 \rangle_H$~\cite{Bodwin:2006dn}.
The generalized Gremm-Kapustin relation made it possible for one 
to resum a class of relativistic corrections of color-singlet
contributions to all orders in $v$~\cite{Bodwin:2006dn}.
The resummation method was applied to improve the NRQCD factorization
formula for the electromagnetic decay of the $J/\psi$.
They determined the parameters of the Cornell potential
model and the normalization of the quarkonium wave function
by making use of the radial mass splittings of the $S$-wave states and
electromagnetic decay widths of the $S$-wave states.
As a result, the LO and order-$v^2$ NRQCD matrix elements of the $S$-wave
quarkonium $H$ were determined~\cite{Bodwin:2006dn}. 
In an updated analysis, they reported
$\langle \mathcal{O}_1 \rangle_{J/\psi}= 0.440^{+0.067}_{-0.055}$ GeV$^3$
and $\langle \bm{q}^2 \rangle_{J/\psi}= 0.441^{+0.140}_{-0.140}$ GeV$^2$~\cite{Bodwin:2007fz}.
This result corresponds to 
$\langle v^2 \rangle_{J/\psi}=0.225^{+0.106}_{-0.088}$, which
is consistent with the naive estimate $v^2 \approx 0.3$ within errors.
This is the most reliable determination of the color-singlet NRQCD matrix
elements for $J/\psi$ available so far~\cite{Bodwin:2007fz}.

The method of resummation~\cite{Bodwin:2006dn} has been applied to
resum a class of the relativistic corrections
to the $e^+ e^- \to J/\psi + \eta_c$~\cite{Bodwin:2006ke}
in combination with the QCD NLO
corrections in Ref.~\cite{Zhang:2005cha}.
In the following updated study in Ref.~\cite{Bodwin:2007ga},
the authors reported the prediction
$\sigma[e^+ e^-\to J/\psi + \eta_c]= 17.6^{+7.8}_{-6.3}$ fb,
which agrees with the experiments within errors.
For more detailed descriptions of estimating theoretical
uncertainties, we refer the readers to Ref.~\cite{Bodwin:2007ga}.
Therefore, in spite of large error bars, it seems fair to conclude
that the long-standing puzzle
between theory and experiments for the $e^+ e^- \to J/\psi+ \eta_c$ 
has been resolved~\cite{Bodwin:2007ga}.

Independently of the theoretical analysis in Ref.~\cite{Bodwin:2007ga},
a fixed-order relativistic corrections 
of relative-order $v^2$ to the cross section for
$e^+ e^- \to J/\psi + \eta_c$ is given in Ref.~\cite{He:2007te},
where the nonperturbative factors
$\langle \mathcal{O}_1 \rangle_{J/\psi}$ and
$\langle \bm{q}^2 \rangle_{J/\psi}$ are determined by making
use of the measured values for 
$\Gamma[J/\psi\to e^+ e^-]$, $\Gamma[\eta_c\to \gamma\gamma]$,
and $\Gamma[J/\psi \to 3g]$ with the assumption 
$\langle \mathcal{O}_1 \rangle_{J/\psi}%
\langle \bm{q}^2 \rangle_{J/\psi}=%
\langle \mathcal{O}_1 \rangle_{\eta_c}%
\langle \bm{q}^2 \rangle_{\eta_c}$. They obtained
the cross section $20.04$ fb with
$\langle \mathcal{O}_1 \rangle_{J/\psi}= 0.573$ GeV$^3$
and $\langle \bm{q}^2 \rangle_{J/\psi}= 0.202$ GeV$^2$~\cite{He:2007te}.
The values for the nonperturbative factors yields 
$\langle v^2 \rangle_{J/\psi}\approx 0.09$. It is smaller by about 
a factor of 2 than that obtained
in Ref.~\cite{Bodwin:2007fz}, which is consistent with 
the naive estimate $v^2\approx 0.3$.
We notice that the short-distance coefficient of $\Gamma[J/\psi\to 3g]$ at
relative-order $v^2$ is very large. This restricts the fit in
Ref.~\cite{He:2007te} to have a small value for
$\langle v^2 \rangle_{J/\psi}$~\cite{Bodwin:2007fz}.
Further applications to other exclusive double-charmonium production
processes are also available in Refs.~\cite{Zhang,Zhang:2008gp}.
\section{Inclusive charm production in bottomonia decays\label{sec:bottom}}
As described in Sec.~\ref{sec:intro}, the earliest calculations of
the widths of $P$-wave quarkonium states using perturbative QCD were
plagued with infrared divergences~\cite{Barbieri:1976fp,Barbieri:1980yp}.
In 1992, Bodwin, Braaten, and Lepage showed that the infrared divergences
could be absorbed into the probability for the $Q\bar{Q}$ pair to be at
the same point in a color-octet state~\cite{Bodwin:1992ye}.
Based on the NRQCD factorization formalism, one can now carry out
rigorous calculations of inclusive charm production from $\chi_{bJ}$ decays.
On the experimental side, $\chi_{bJ}(1P)$ and
$\chi_{bJ}(2P)$ have been discovered. The only properties of these states
that have been measured thus far are their masses and their radiative
branching fractions into $\Upsilon(nS)$.
The total widths of the $\chi_{bJ}(nP)$ states have not 
been measured. Recent runs of the CLEO experiment at the $\Upsilon(2S)$
and $\Upsilon(3S)$ resonances have provided new data on the $\chi_{bJ}(1P)$
and $\chi_{bJ}(2P)$ states. The $B$-factory experiments BABAR and Belle can
study the $\chi_{bJ}(nP)$ states by using data samples of $\Upsilon(2S)$
and $\Upsilon(3S)$ provided by initial-state radiation. The Belle
experiment has also accumulated data by running directly on the
$\Upsilon(3S)$ state. Therefore, it is necessary for theorists
to present quantitative predictions for the $P$-wave
bottomonium states that can be compared with forthcoming data.

The widths of all four states in a $P$-wave multiplet can be
calculated by using the NRQCD factorization formula,
once the two nonperturbative factors $\langle \mathcal{O}_1\rangle_{\chi_b}$ 
and $\langle \mathcal{O}_8\rangle_{\chi_b}$ have
been determined. In Ref.~\cite{Bodwin:2007zf}, the authors computed the
inclusive charm production rate in the $\chi_{bJ}$
decay $\chi_{bJ}\to c+X$. They also provided the momentum distributions
of charmed hadrons produced in the process by making use of
a parametrization for the charm fragmentation function
fit to the data from $e^+e^-$ collisions~\cite{Seuster:2005tr}.
The predictions are
dependent on the ratio 
$\rho_8\equiv m_b^2\langle \mathcal{O}_8\rangle_{\chi_b}/%
\langle \mathcal{O}_1\rangle_{\chi_b}$ that can be determined
phenomenologically.
Recently, the CLEO Collaboration measured
the total production rates for the charm in $\chi_{bJ}$  decays and
the branching fractions for $\chi_{bJ}(1P,2P)\to D^0 X$ with
$p_{D^0}>2.5$ GeV~\cite{Briere:2008cv}. The CLEO measurement indeed
determined $\rho_8(1P) = 0.160^{+0.071}_{-0.047}$ and 
$\rho_8(2P)= 0.074^{+0.010}_{-0.008}$~\cite{Briere:2008cv}.
The joint fit for $1P$ and $2P$ gives
$\rho_8 = 0.086^{+0.009}_{-0.013}$~\cite{Briere:2008cv}.

As shown in Ref.~\cite{Bodwin:2007zf}, the branching fraction
$\textrm{Br}[\chi_{bJ}\to c+X]$ must be particularly large for
the $J=1$ state. The reason is that the $\chi_{b1}\to c\bar{c}g$ 
is one of the LO contributions $\chi_{b1}\to q\bar{q}g$ to
hadronic decay of $\chi_{b1}$, while the mode $\chi_{bJ}\to c\bar{c}g$
is suppressed by order $\alpha_s$ compared to
the leading contribution $\chi_{bJ}\to gg$ 
for hadronic decays of $J=0$ and 2 states.
Note that $\chi_{b1}\to gg$ is forbidden by Yang's theorem.
The measurement of the CLEO Collaboration 
also confirmed the
large branching fractions of about $25$\% for both
 $\chi_{b1}(1P)$ and
 $\chi_{b1}(2P)$~\cite{Briere:2008cv}, which is consistent with
 the prediction in Ref.~\cite{Bodwin:2007zf} within errors.

Theoretical studies on the inclusive charm production has been
extended to the $S$-wave bottomonium states in
Refs.~\cite{Kang:2007uv,Chung:2008yf,Zhang:2008pr,Hao:2007rb}.
\section{Conclusions\label{sec:sum}}
The discrepancy between the theoretical prediction and the CDF measurements
at Run I and Run II of the Fermilab Tevatron for the prompt $S$-wave
spin-triplet charmonium produced at large $p_T$ is embarrassing.
To make the problem more complicated, there is a big
discrepancy between the CDF data of Run I and those of Run II that
has not been clearly understood, yet.
In order to find any other contributions that may modify the theoretical
prediction toward experimental results, efforts have been made
in various manners. A few sources which contribute to the longitudinal 
polarization for the prompt $J/\psi$ and $\psi(2S)$ are found. However,
the complete analysis which includes all of these contributions 
and includes all the feeddowns that are contained in the data
is not available, yet. In order to make a reliable prediction for the
polarization with these new contributions, more accurate determination
of various NRQCD matrix elements for the quarkonium production must
be completed, which may not be an easy task.

The puzzle of the cross section for 
$e^+e^-\to J/\psi+\eta_c$ at the $B$ factories has
spurred rapid developments in heavy-quarkonium phenomenology.
It led the first reliable estimate of the
relative-order-$v^2$ color-singlet NRQCD matrix element
for the $J/\psi$,
that was uncertain even in the sign.
A new method to resum relativistic corrections to a class of 
color-singlet contributions has been developed to determine the
color-singlet
NRQCD matrix elements for the $S$-wave heavy quarkonia more accurately.
A recent computation of order-$\alpha_s$ corrections to the
quarkonium electromagnetic current to all orders in $v$ revealed
that the velocity expansion for approximate $J/\psi$ operator
matrix elements converges very rapidly~\cite{Bodwin:2008vp}.
This supports the reliability of the resummation method.
Being applied to the $e^+e^-\to J/\psi+\eta_c$ process, the resummation
method for the relativistic corrections has resolved the
puzzle in combination with large QCD corrections.
The puzzle also motivated the proofs of the factorization
theorems for exclusive two-body charmonium production in $B$-meson decay
and $e^+e^-$ annihilation to all orders in perturbation theory in quantum
chromodynamics~\cite{Bodwin:2008nf}.

Resolution of the infrared-singularity problem in $P$-wave heavy-quarkonium
decay into hadrons was the first successful theoretical application
of the NRQCD factorization.
However, the process has never been thoroughly investigated experimentally.
A new measurement by the CLEO Collaboration of the inclusive charm
production rate in $P$-wave bottomonium decay has provided useful
information. The measurement helped us determine the ratio for the
color-octet NRQCD matrix element to the color-singlet matrix element
for the $P$-wave states accurately. 
On-going analysis by the CLEO Collaboration on the charm-hadron momentum 
distribution may reveal the color-octet
mechanism in $P$-wave bottomonium decay more clearly.

\end{document}